\documentclass[preprint2]{emulateapj}

\shorttitle{Proper-Motion of NE Forward Shock of SN~1006}
\shortauthors{S. Katsuda et al.}

\begin{document}

\title{The First X-Ray Proper-Motion Measurements of the Forward Shock
  in the Northeastern Limb of SN~1006}

\author{Satoru Katsuda\altaffilmark{1,2}, Robert
  Petre\altaffilmark{1}, Knox S. Long\altaffilmark{3}, Stephen
  P. Reynolds\altaffilmark{4},  P. Frank Winkler\altaffilmark{5},
  Koji Mori\altaffilmark{6}, and Hiroshi Tsunemi\altaffilmark{2}
}

\email{Satoru.Katsuda@nasa.gov}

\altaffiltext{1}{NASA Goddard Space Flight Center, Greenbelt, MD
   20771, U.S.A.} 

\altaffiltext{2}{Department of Earth and Space Science, Graduate School
of Science, Osaka University,\\ 1-1 Machikaneyama, Toyonaka, Osaka,
560-0043, Japan}

\altaffiltext{3}{Space Telescope Science Institute, 3700 San Martin
  Dr., Baltimore, MD 21218, U.S.A.} 

\altaffiltext{4}{Physics Department, North Carolina State University, Raleigh, North Carolina 27695} 

\altaffiltext{5}{Department of Physics, Middlebury College,
  Middlebury, VT 05753}

\altaffiltext{6}{Department of Applied Physics, Faculty of Engineering,
University of Miyazaki, 1-1 Gakuen Kibana-dai Nishi, Miyazaki, 889-2192,
Japan}

\begin{abstract}

We report on the first X-ray proper-motion measurements of the
nonthermally-dominated forward shock in the northeastern limb of
SN~1006, based on two {\it Chandra} observations taken in 2000 and
2008.  We find that the proper motion of the forward shock is about
0.48$^{\prime\prime}$\,yr$^{-1}$ and does not vary 
around the rim within the $\sim$10\% measurement uncertainties.  The
proper motion measured is consistent with that determined by the
previous radio observations.  The mean expansion index of the forward
shock is calculated to be $\sim$0.54 which matches the value expected
based on an evolutionary model of a Type Ia supernova with either a
power-law or an exponential ejecta density profile. Assuming pressure
equilibrium around the periphery from the thermally-dominated
northwestern rim to the nonthermally-dominated northeastern rim, we
estimate the ambient density to the northeast of SN~1006 to be
$\sim$0.085\,cm$^{-3}$.
\end{abstract}

\section{Introduction}

SN~1006 is a Galactic shell-type supernova remnant (SNR) originating 
from a Type Ia SN explosion.  In X-rays, its northeastern (NE) and
southwestern (SW) limbs are dominated by nonthermal synchrotron 
emission.  These provided the first solid evidence that shells of SNRs
are efficient accelerators of cosmic rays (Koyama et al.\ 1995;
Winkler \& Long 1997b).  
After the discovery of the nonthermal synchrotron emission from the NE
and SW limbs of SN~1006, TeV $\gamma$-ray emission had been expected to be
detected, but has not yet been established\footnote{The HESS team
  recently reported the detection of TeV emission from SN~1006 (De 
  Naurois et al.\ 2008), which is consistent with the upper limit
  previously determined (Aharonian et al.\ 2005).  Although their
  result seems reliable, the result is not yet published and they
  have not yet shown the TeV spectrum.  Therefore, we do not take
  account of the detection in this paper.}
(Aharonian et al.\ 2005). TeV
emission associated with SNR shocks can arise from either of two processes: 
inverse Compton scattering of cosmic microwave
background photons by relativistic electrons (the leptonic process),
or $\pi^0$-decay emission resulting from accelerated protons
interacting with the ambient medium (the hadronic process).  The TeV flux
from the hadronic process strongly depends on the ambient density
(Ksenofontov et al.\ 2005).  Thus, measuring the ambient density
is important for understanding the origin of any TeV $\gamma$-ray
emission detected from SN~1006.  

The ambient density around the thermally emitting NW and SE rim can be
measured based on spectral modelling of the thermally-dominated
spectra.  On the other hand, for the NE and SW rims 
where nonthermal emission dominates, it is quite difficult to measure
the ambient density from spectral analyses.  We may, however, infer
the ambient density around nonthermally-dominated rims indirectly by assuming
constancy of $n_0 v_\mathrm{s}^2$, where $n_0$ is the ambient density
and $v_\mathrm{s}$ is the shock velocity (i.e., a uniform pressure
around the periphery from the thermally-dominated NW rim to the
nonthermally-dominated NE rim).  In this context, measuring the 
proper motion of the forward shock is important for estimating the
ambient density around nonthermally-dominated rims of SNRs and
constraining the origin of TeV $\gamma$-ray emission. 

So far, proper-motion measurements of the SN~1006 have been performed by
optical and radio observations.  Optical proper-motion of 
the bright NW H$\alpha$ filament   was first measured by Long, Blair,
\& van den Bergh (1988) and later by Winkler, Gupta, \& Long (2003),
who precisely determined the proper motion to be
0.280$^{\prime\prime}\pm$0.008$^{\prime\prime}$\,yr$^{-1}$ along the
entire length of the NW filament.  The radio observations measured the 
mean proper motion for the entire rim of the remnant to be
0.44$^{\prime\prime}\pm$0.13$^{\prime\prime}$\,yr$^{-1}$ (Moffett et
al.\ 1993).  Here, we report on the first X-ray proper-motion
measurements of NE rim of SN~1006, based on two {\it Chandra}
observations  separated by about 8 yr.

\section{Observations}

We observed the NE rim of SN~1006 twice: first on 2000 July 10
(ObsID.\ 732; PI: Long) and a second time on 2008 June 24
(ObsID.\ 9107; PI: Petre).  The second observation was
specifically intended to allow a proper motion measurement; we
requested the same pointing direction, roll angle and exposure time as
the previous observation.  The time difference between the two
observations was 8.04 yr.  We start our analysis from level 2 event
files processed with calibration data files in CALDB ver.\ 3.4.0 for
ObsID.\ 732, ver.\ 3.4.5 for ObsID.\ 9107.  Fortunately, we see no
background flares in the data from ObsID.\ 9107 so that we reject no
data from the level 2 event file for this data set.  On the other
hand, we exclude high-background periods for data from ObsID.\ 732.
The resulting net exposure times for ObsID.\ 732 and 9107 are 55.3\,ks
and 68.7\,ks, respectively.

\section{Analysis and Results}

\begin{figure}
\includegraphics[angle=0,scale=0.45]{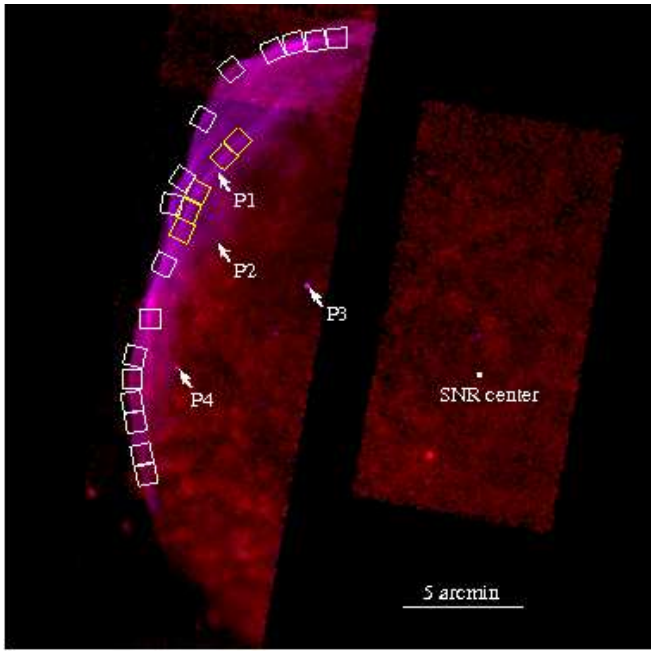}\hspace{1cm}
\caption{{\it Chandra} two-color image after vignetting effects are
  corrected.  Red and blue represent 0.3--1.0\,keV and
  1.0--8.0\,keV X-rays, respectively.  The image is binned by
  1.97$^{\prime\prime}$ and has been smoothed by a Gaussian kernel of
  $\sigma = 5.90^{\prime\prime}$.  The intensity scale is square
  root.  We measure proper motions in the white and yellow box regions
  shown in this figure.  The four reference stars are indicated as
  arrows (from P1 to P4). 
} 
\label{fig:image}
\end{figure}

Figure~\ref{fig:image} shows a two-color image obtained from the 2008
observation, after exposure correction to remove vignetting effects.  
Red and blue correspond to a low-energy (0.3--1.0\,keV)
and a high-energy (1.0--8.0\,keV) band, respectively.  The nonthermal
forward shocks appear as bluish filamentary structures along the edge
of the remnant.  In the following analysis, we measure proper motions
of various portions of the forward shocks by comparing the images
obtained in the two epochs.  We use an energy range of 1.0--8.0\,keV
to measure proper motions of the forward shocks.   

Although the absolute astrometric accuracy for the Advanced CCD Imaging
Spectrometer reported by the {\it Chandra} calibration
team\footnote{http://cxc.harvard.edu/cal/} reached a good level of 90\%
uncertainty = 0.6$^{\prime\prime}$, we further 
align the images taken in different two epochs by aligning positions
of four point sources which are visible in the FOV.  These four 
sources are marked as P1--P4 in Fig.~\ref{fig:image}.  
We determine their positions at each epoch, using  {\tt
  wavdetect} software included in CIAO ver.\ 4.0.  Three of them 
(i.e., P1, P3, and P4 marked in Fig.~\ref{fig:image}) have optical
counterparts, based on the Naval Observatory Merged Astrometric
Dataset (NOMAD) catalog (Zacharias et al.\ 2005). The proper motions
of the three sources are negligible, at less than
$\sim$0.005$^{\prime\prime}$yr$^{-1}$ (Note that the P3 source is
identified as a QSO; Winkler \& Long 1997a).  Although we cannot
find an optical counterpart for the P2 source, the X-ray positions
determined for the two epochs are consistent within the errors.  Thus,
we assume that the proper motion of the P2 source is also negligible. 
Then, we align the coordinates of the images so that the four
point sources have the same coordinates in the two epochs, by using
{\tt ccmap} and {\tt ccsetwcs} in IRAF IMCOORDS.  In the alignment
procedure, four parameters for the second-epoch image are allowed to
vary: x and y shifts, rotation, and pixel scale.
The best-fit parameters show very small differences; both x and y
shifts are less than 0.2$^{\prime\prime}$, rotation is less than
50$^{\prime\prime}$, and no modification is required for 
the pixel scale.  After the alignment, we find the rms residuals  for the
positions of the four point sources between the two epochs 
to be $\sim$0.14$^{\prime\prime}$.  We take this value  as a
systematic uncertainty in our proper-motion measurements.  

\begin{figure}
\includegraphics[angle=0,scale=0.45]{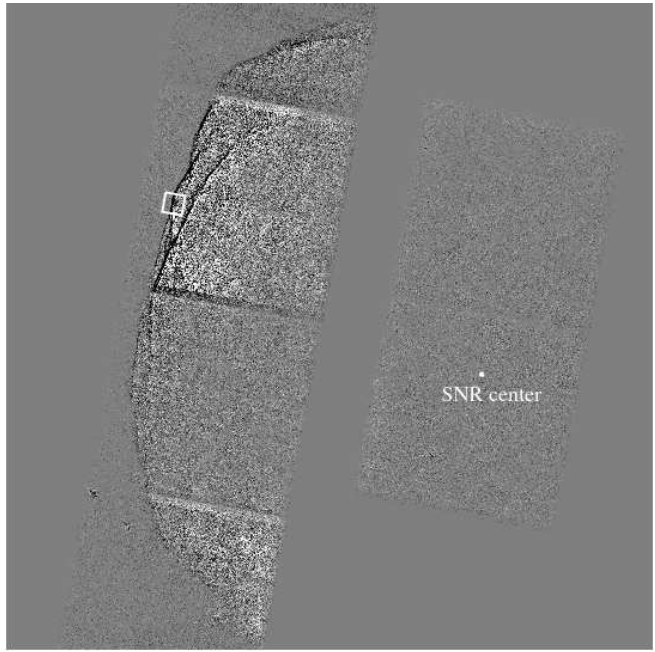}
\caption{Difference image between the first- and second- epoch in
  1.0--8.0\,keV band.  The image is binned by 0.492$^{\prime\prime}$
  and has been smoothed using a Gaussian kernel of $\sigma =
  1.467^{\prime\prime}$.   The intensity is linearly scaled
  from -0.5 to +0.5 counts\,bin$^{-1}$.  Example one-dimensional
  profiles from the box area are shown in Fig.~\ref{fig:onedim}.
}
\label{fig:diff}
\end{figure}

Figure~\ref{fig:diff} shows the difference image between the two
epochs in the 1.0--8.0\,keV band, after being exposure-corrected,
registered and normalized to match the count rates in the two epochs.
The narrow black lines along the periphery in this difference image
clearly indicate the motion of the forward shock. 
In order to measure the proper motions accurately, we generate
one-dimensional X-ray profiles along the apparent direction of
motion---extracted from 21 areas (white and yellow boxes in
Fig.~\ref{fig:image}) along the forward shock.  Yellow boxes indicate
inner shocks where there are essentially two shocks at the same
azimuth angle (presumably a projection effect).  In each case we have
extracted the profiles in a direction perpendicular to the forward
shock, as determined from the local tangent to the rim taken at each
of the extraction boxes.  Varying this tangent direction by several
degrees does not significantly affect the results. 

\begin{figure}
\includegraphics[angle=0,scale=0.33]{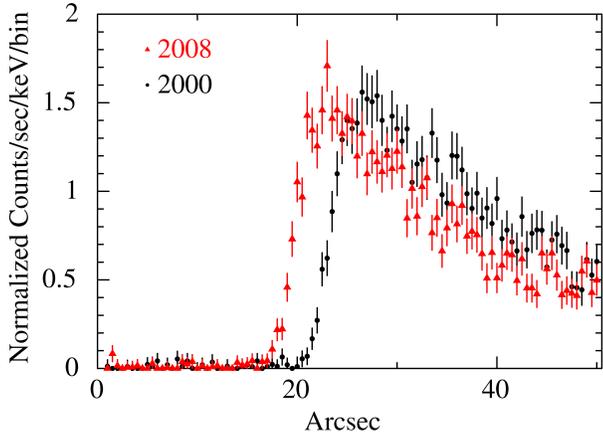}
\caption{Example one-dimensional profiles extracted from the area
  shown in Fig.~\ref{fig:diff}. Data points with triangles and circles
  are responsible for 2008 and 2000 observations, respectively.} 
\label{fig:onedim}
\end{figure}

To calculate the expansion rates and expansion indices, we estimated
the center for the SNR shell by fitting a circle to the outer
surface-brightness contour of the whole remnant based on a {\it ROSAT}
HRI image: RA $ = $ 15$^\mathrm{h}$02$^\mathrm{m}$54$^\mathrm{s}$.9 
(J2000), DEC $= -41^{\circ}56^{\prime}08.9^{\prime\prime}$ (J2000),
shown in Fig.~\ref{fig:image} and Fig.~\ref{fig:diff}.  This center
lies within $\sim 40\arcsec$\ of centers for the entire 
SN~1006 shell estimated from radio (Roger et al.\ 1988), X-ray
(estimated by eye on a {\it ROSAT} PSPC image: Willingale et al.\
(1996), and optical (Winkler, Gupta, \& Long
2003) images\footnote{Winkler, Gupta, \& Long (2003) also give a
  center of curvature for bright NW optical filaments where they
  measured the proper motions; this position is offset
  $\sim5$\arcmin~SE from the shell center.}.  The numbers of photons
from the two epochs are equalized in each area by scaling.  Each
one-dimensional X-ray profile is binned by 0.492$^{\prime\prime}$.  An
example one-dimensional profile is shown in Fig.~\ref{fig:onedim}, in
which we can clearly see the motion of the shock.  Then, shifting the
first-epoch profile and calculating the $\chi^2$-values from the
difference between the two profiles at each shift position, we search
for the best-matched shift position for each area.  The reduced 
$\chi^2$-values, which are nearly equal to $\chi^2$-value divided by
the number of bins in each area (here 69), range from 0.5 to 1.2 at
the best-matched shift positions.  In this way, we obtain a proper
motion for each area.  A detailed explanation of our method of
proper-motion measurements can be found in Katsuda et al.\ (2008).   

\begin{figure}
\includegraphics[angle=0,scale=0.65]{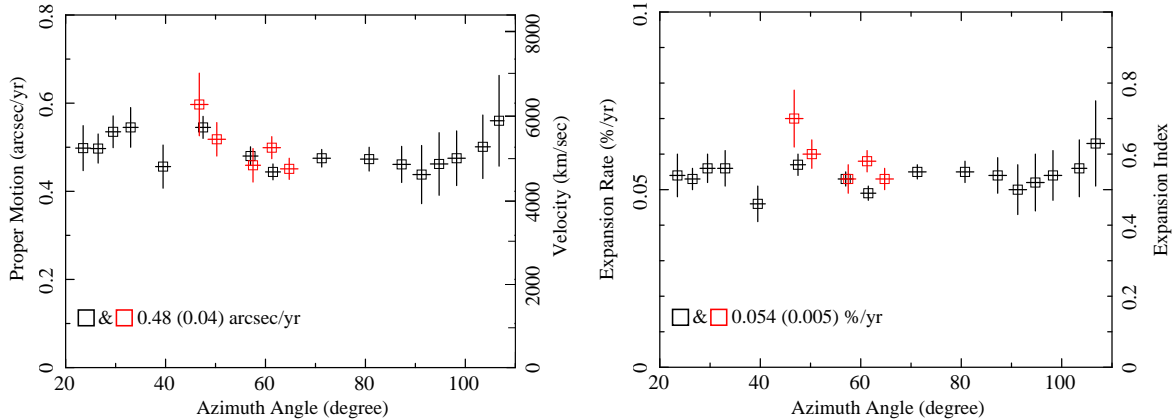}
\caption{{\it Left}: Proper motion (left axis) and velocitiy for a
  distance of 2.2\,kpc (right axis) as a function of azimuthal angle
  (counterclockwise from the north).  The black and red data points
  are obtained from the white and yellow boxes in Fig.~\ref{fig:image},
  respectively.  The weighted mean proper motion (with rms) is shown
  in the panel. {\it Right}: Same as left but for 
  expansion rate (left axis) and expansion index (right axis).}  
\label{fig:results}
\end{figure}

Figure~\ref{fig:results} {\it left} shows both proper motions and
velocities (for a distance of 2.2\,kpc; Winkler, Gupta, \& Long 2003)
for all the areas as a function of azimuthal angle, and
Fig.~\ref{fig:results} {\it right} shows both expansion rates and
expansion indices, $m$; $R \varpropto t^m$ and $m$ is derived from
(the expansion rate) $\times$ 
(the age of the remnant, 998 yr for observations used here).  The data
points with black and red are obtained from the white and yellow boxes in
Fig.~\ref{fig:image}, respectively.  The errors quoted are statistical
1$\sigma$ uncertainties.  We find that the proper 
motion around the forward shocks is almost invariant.  Also notable is
that there is no significant velocity difference between the inner
shock and the outer shock at the azimuth angle around
50$^{\circ}$--70$^{\circ}$.  The weighted mean value of the proper
motions of the forward shocks is calculated to be
$\sim$0.48$^{\prime\prime}$\,yr$^{-1}$ with the rms deviation of
$\sim$0.04$^{\prime\prime}$\,yr$^{-1}$.  The expansion rates and the
expansion indices are also nearly constant at about
0.054\,\%\,yr$^{-1}$ and 0.54, respectively.  The statistical 
error is about 10\% ($\sim$0.05$^{\prime\prime}$\,yr$^{-1}$) for each
data point, so that it is about three times greater than the
systematic uncertainty of  
0.017$^{\prime\prime}$\,yr$^{-1}$ (= 0.14$^{\prime\prime}$/8.04 yr).
Thus, the current expansion index of the NE rim is considerably less
than if it were in a free expansion ($m = 1$), but significantly higher 
than the value for an interstellar medium (ISM)-dominated
adiabatically expanding SNR ($m = 0.4$). 

\section{Discussion}

This paper presents the first X-ray proper-motion measurement of the 
forward shock along the NE rim of SN~1006.  The measured proper motion 
of $\sim$0.48$^{\prime\prime}\pm$0.04$^{\prime\prime}$\,yr$^{-1}$\ 
does not vary significantly with azimuth.  This value is consistent
with the mean expansion rate for the entire SN~1006 shell measured by
radio observations, 
0.44$^{\prime\prime}\pm$0.13$^{\prime\prime}$\,yr$^{-1}$\ (Moffett 
et al.\ 1993).
For other young SNRs, there have been problems reconciling expansion
index measurements from X-ray and radio wavelengths, with the X-ray
rate seeming to be larger in Cas A, Kepler's SNR and Tycho's SNR as
noted by e.g., Hwang et al.\ (2002) or DeLaney et al.\
(2004).  Table~\ref{tab:index} summarizes the radio and X-ray
expansion indices for these SNRs.  We note that recent {\it
  Chandra}-based X-ray proper-motion measurements have reduced the
expansion index for Kepler, more in line with the radio value.  Cas~A 
is the only SNR whose X-ray expansion index as measured with {\it
  Chandra} clearly differs from its radio value.  Tycho's SNR has yet
to be remeasured with {\it Chandra}, but should be as soon as
possible. 

\begin{deluxetable}{lccccccccc}
\tabletypesize{\scriptsize}

\tablecaption{Expansion Index Measurements for Young SNRs}
\tablewidth{0pt}
\tablehead{
\colhead{SNR name}& &&&&\colhead{Radio} &&&&\colhead{X-rays}  
}
\startdata
Cas~A&&&&&0.35$^{1}$&&&&0.73$^{1,2,3}$\\
Tycho&&&&&0.47$^{4}$&&&&0.71$^{5}$\\
Kepler&&&&&0.5$^{6}$&&&&0.93$^{7} \rightarrow$ 0.35--0.8$^{8,9}$ \\
SN~1006&&&&&0.48$^{10}$&&&&0.54$^{11}$
\enddata

\tablecomments{1: Koralesky et al.\ (1998), 2: Vink et al.\ (1998), 3:
  DeLaney \& Ludnick (2003),
  4: Reynoso et al.\ (1997), 5: Hughes (2000), 6: Dickel et al.\ (1988),
  7: Hughes (1999), 8: Katsuda et al.\ (2008), 9: Vink (2008), 10:
  Moffett et al.\ (1993), 11: This work}
\label{tab:index}
\end{deluxetable}

We here estimate the ambient density around the NE rim of SN~1006 and
discuss its implication for TeV $\gamma$-ray emission.  Along the NW 
rim, the proper motion of the forward shock is well determined to be
0.280$^{\prime\prime}\pm$0.008$^{\prime\prime}$ \,yr$^{-1}$ by 
optical observations (Long, Blair, \& van den Bergh 1988; Winkler,
Gupta, \& Long 2003).  The ratio of the forward shock velocities 
between the NE rim and the NW rim is estimated to
be $\sim$1.7.  The ambient density ahead of the NW rim is measured to be
0.15--0.25\,cm$^{-3}$ by X-ray spectroscopy (Long et al.\ 2003; Acero
et al.\ 2007) or 0.25--0.4\,cm$^{-3}$ by optical spectroscopy (Raymond
et al.\ 2007).  We here take 0.25$^{+0.15}_{-0.10}$\,cm$^{-3}$ as a
conservative ambient density value.  Then, assuming pressure
equilibrium around the periphery from the NW rim to the NE rim (i.e.,
constant $n_0 v_\mathrm{s}^2$), we estimate the ambient density ahead
of the NE rim to be 0.085$^{+0.055}_{-0.035}$\,cm$^{-3}$.  This is
consistent with the upper limit, 0.1\,cm$^{-3}$, constrained by an
upper limit of TeV $\gamma$-ray flux obtained by HESS (Aharonian et
al.\ 2005) combined with a $\gamma$-ray emission model (Ksenofontov et
al.\ 2005).  

We derived the mean expansion index of the NE forward shock to be
$\sim$0.54. Dwarkadas \& Chevalier (1998) examined the evolution of 
Type Ia SNe for both power-law and exponential density profiles of
ejecta.  From their calculations, the evolution of the forward shock
is scaled by the time, $t^{\prime} = 248
(n_0/1\,\mathrm{cm^{-3}})^{-1/3}$ yr.  The age of SN1006, 998 
yr, corresponds to 1.48--2.08 $t^{\prime}$, depending on a range of 
the ambient density of 0.05--0.14\,cm$^{-3}$.  Then, the expansion index is
expected to be 0.57--0.53 (power-law) or 0.54--0.51 (exponential).
Thus, we find that the measured value is in good agreement with that
expected in the Type Ia SNe model with either a power-law or
an exponential ejecta density profile. 

We note that there are a number of thermally-dominated knotty features
in the remnant (see, Fig.\ref{fig:image}).  The origin of these
features are suggested to be either fingers formed by Rayleigh-Taylor
instabilities acting at the contact discontinuity between the swept-up
ISM and the ejecta (Long et al.\ 2003; Cassam-Chena\"i et al.\ 2007)
or ejecta bullets (Long et al.\ 2003; Vink et al.\ 2003).  It is also
interesting to measure proper motions for these features, which is
beyond the scope of this paper and remains as our future work.

\section{Conclusion}

We, for the first time, performed X-ray proper-motion measurements of
the forward shock in the NE limb of SN~1006.  The proper motion of the
forward shock is derived to be $\sim$0.48$^{\prime\prime}$\,yr$^{-1}$.
This is consistent with the results from radio observations.
The expansion index of the forward shock is calculated to be
$\sim$0.54, which supports that SN~1006 is a remnant of a Type Ia
supernova with either a power-law or an exponential ejecta density
profile.  We estimate the ambient density to the NE of SN~1006 to be
$\sim$0.085\,cm$^{-3}$.  Our derived ambient density of
0.085\,cm$^{-3}$ is below the limit of 0.1\,cm$^{-3}$ obtained by 
Ksenofontov et al.\ (2005) in modeling hadronic and leptonic TeV
$\gamma$-ray emission subject to the HESS upper limit (Aharonian et
al.\ 2005).

\acknowledgments

We acknowledge helpful scientific discussions with Una Hwang and
Hiroya Yamaguchi.
S.K.\ is supported by a JSPS Research
Fellowship for Young Scientists.  S.K.\ is also supported in part by
the NASA grant under the contract NNG06EO90A. 
P.F.W.\ acknowledges the support of the NSF through grant AST 03-07613.

\end{document}